\documentclass[%
 aip,apl,
 amsmath,amssymb,
 reprint,%
]{revtex4-1}
\usepackage{graphicx}
\usepackage[utf8]{inputenc}
\usepackage[T1]{fontenc}
\usepackage{etoolbox}
\usepackage{lineno}
\usepackage{xcolor}
\newcommand\mycolor[1]{\textcolor{black}{#1}}
\newcommand\mycolorbis[1]{\textcolor{black}{#1}}
\usepackage{comment}
\usepackage{makecell}
\newcommand{\SAP}{Dipartimento di Fisica - Sapienza Universit\`{a} di Roma, Piazzale Aldo Moro 2, 00185, Roma - Italy}

\newcommand{\INFNRM}{INFN - Sezione di Roma, Piazzale Aldo Moro 2, 00185, Roma - Italy}
\newcommand{\INFNFE}{INFN - Sezione di Ferrara, via Saragat,1 44121, Ferrara - Italy}
\newcommand{\UNIFE}{Dipartimento di Fisica e Scienze della Terra, Università di Ferrara, Via Saragat 1, 44100, Ferrara, Italy}

\newcommand{\NEEL}{Univ. Grenoble Alpes, CNRS, Grenoble INP, Institut N\'eel, 38000 Grenoble, France}

\usepackage{comment} % to remove togheter with comments

\begin{document}%\linenumbers
\title{Germanium target sensed by phonon-mediated kinetic inductance detectors}
\author{D.~Delicato}\email{daniele.delicato@roma1.infn.it}\affiliation{\NEEL}\affiliation{\SAP}\affiliation{\INFNRM}
\author{D.~Angelone}\affiliation{\SAP}\affiliation{\INFNRM}
\author{L.~Bandiera}\affiliation{\INFNFE}
\author{M.~Calvo}\affiliation{\NEEL}
\author{M.~Cappelli}\affiliation{\SAP}\affiliation{\INFNRM}
\author{U.~Chowdhury}\affiliation{\NEEL}
\author{G.~Del~Castello}\affiliation{\SAP}\affiliation{\INFNRM}
\author{M.~Folcarelli}\affiliation{\SAP}\affiliation{\INFNRM}
\author{M.~del~Gallo~Roccagiovine}\affiliation{\SAP}\affiliation{\INFNRM}
\author{V.~Guidi}\affiliation{\UNIFE}\affiliation{\INFNFE}
\author{G.~L.~Pesce}\affiliation{\SAP}\affiliation{\INFNRM}
\author{M.~Romagnoni}\affiliation{\INFNFE}
\author{A.~Cruciani}\affiliation{\INFNRM}
\author{A.~Mazzolari}\affiliation{\UNIFE}\affiliation{\INFNFE}
\author{A.~Monfardini}\affiliation{\NEEL}
\author{M.~Vignati}\affiliation{\SAP}\affiliation{\INFNRM}

\begin{abstract}
Cryogenic phonon detectors are adopted in experiments searching for dark matter interactions or coherent elastic neutrino-nucleus scattering, thanks to the low energy threshold they can achieve.
The phonon-mediated sensing of particle interactions in passive silicon absorbers has been demonstrated with Kinetic Inductance Detectors (KIDs).
Targets with neutron number larger than silicon, however, feature higher cross section to neutrinos while multi-target absorbers in dark matter experiments would provide a stronger evidence of a possible signal.
In this work we present the design, fabrication and operation of KIDs coupled to a germanium absorber, achieving phonon-sensing performance comparable to silicon absorbers.
The device introduced in this work is a proof of concept for a scalable neutrino detector and for a multi-target dark matter experiment.

\end{abstract}

\maketitle

%%%%%%%%%%%%%%%%%%%%%%%%%%%%%%%%%%%%%%%%%%%%%%%%%%%%%%%%%%%%%%%%%%%
%\section{INTRODUCTION}
Cryogenic phonon detectors achieve the low energy threshold needed to search for dark matter interactions~\cite{SuperCDMS:2017mbc,CRESST2019,EDELWEISS:2016boq} and neutrino coherent and elastic scattering
off atomic nuclei (CE$\nu$NS)~\cite{Akimov:2017ade,StraussGram,Billard_2017}. 
Such detectors consists of a target crystal, acting as particle absorber, coupled to a phonon sensor.

While detector units with energy thresholds as low as 20 eV~\cite{StraussGram,SuperCDMSCPD} have already been employed,
this is only part of the challenge. In order to observe rare signals, such as those induced by dark matter and neutrino interactions, it is also crucial \mycolor{to develop instruments with high target mass}. Although targets of up to several tens of grams have been realized~\cite{CRESST2019,RICOCHETAG,SuperCDMSCPD}, 
reaching the kg scale would allow experiments searching for weakly interacting massive particles (WIMPs~\cite{Roszkowski_2018}) to probe smaller cross-sections~\cite{appec}, and would enable higher precision in CE$\nu$NS experiments in order to search for new physics \cite{Dodd:1991ni,Barranco:2005yy,Formaggio:2011jt,Dutta:2015nlo,Lindner:2016wff}. 

The BULLKID project developed a cryogenic detector designed to address the mass scaling challenge~\cite{bullkid2022}. The detector consists of a monolithic array of 60 cubic silicon absorbers of 0.34~g each carved out of a single 3" diameter 5~mm thick wafer.
The phonon sensors are Kinetic Inductance Detectors (KIDs), superconducting resonators featuring intrinsic frequency multiplexing, \mycolor{ease and consistency in fabrication}~\cite{Day:2003fk}. 

\mycolor{KIDs are used to detect} millimeter wave~\cite{refId0} and optical radiation~\cite{Mazin2019}: incoming photons absorbed in the metal break Cooper pairs and change its kinetic inductance $L_k$ which converts into a measurable shift of the resonant frequency $f_0$.
In the case of BULLKID, KIDs are instead kept in the dark to minimize light-induced pair breaking and maximize the sensitivity to athermal phonons. 
\mycolor{This phonon mediated detection approach allowed to sample the energy deposited into each silicon absorber with an average baseline resolution of 26~eV}~\cite{bullkid2022}.
While this silicon device demonstrated the potential of the design in the field of dark matter searches~\cite{bullkid2024}, the application to CE$\nu$NS detection would benefit from a different target crystal. 

For a nucleus at rest with $N$ neutrons and $Z$ protons the cross-section of the coherent scattering of a neutrino with energy $E_\nu$ is~\cite{Freedman:1977xn} $$\sigma_{SM} \propto G_F^2 \left[ N - Z\cdot (1 -4\cdot \sin^2{\theta_w}) \right]^2 E_\nu^2$$ 
with $G_F$ being the Fermi constant and $\theta_w$ the Weinberg angle, $\sin^2{\theta_w} \approx 0.238$~\cite{thetaw_lowe}. The cross-section is dominated by the neutron number $N$ thus changing the target material from silicon ($N_{Si} = 14$) to germanium ($N_{Ge}\sim41$) increases the cross-section by a factor $\left( N_{Ge}/ N_{Si} \right)^2 \sim 9$.

A multi-target particle detector would also provide an advantage in the field of dark matter identification: the ability to test the quadratic scaling of the WIMP-nucleon cross section as a function of the atomic number A is a fundamental tool to confirm a positive evidence of WIMP interactions. The comparison of the signal from germanium and silicon absorbers in the same composite array will also allow to reduce systematic errors~\cite{appec}.

In this work we describe the design, fabrication and operation of KIDs on a germanium substrate. The performances are in line with those obtained on silicon, making it a promising target material for future experiments with KIDs.\\

%%%%%%%%%%%%%%%%%%%%%%%%%%%%%%%%%%%%%%%%%%%%%%%%%%%%%%%%%%%%%%%%%%%
%\section{DEVICE DESIGN AND FABRICATION}
\label{design}
\begin{figure}[t]
\centering
\includegraphics[width=0.45\textwidth]{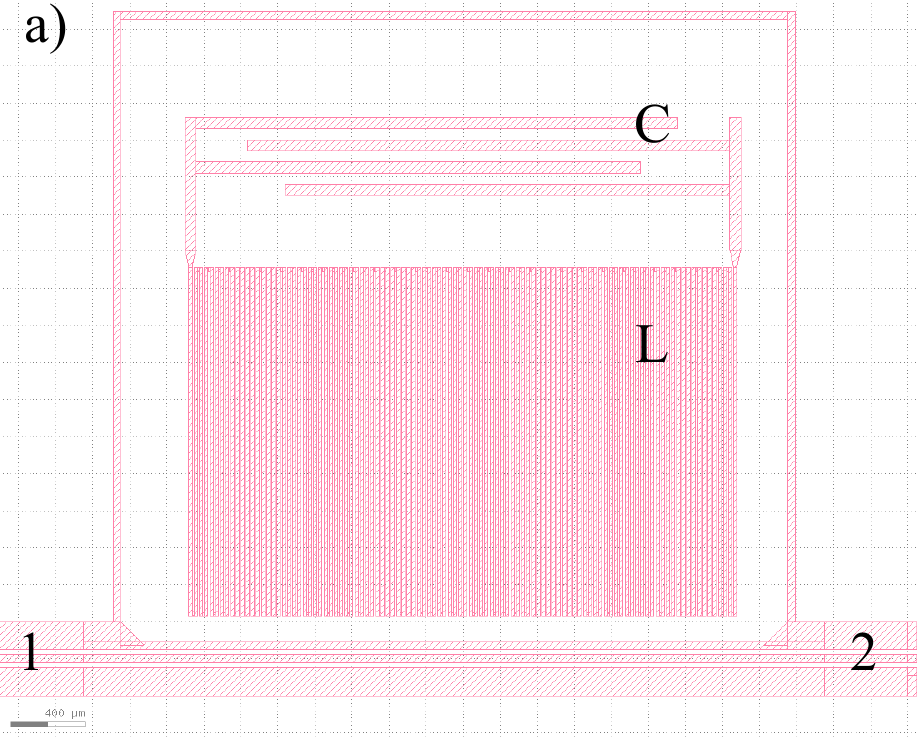}
\includegraphics[width=0.45\textwidth]{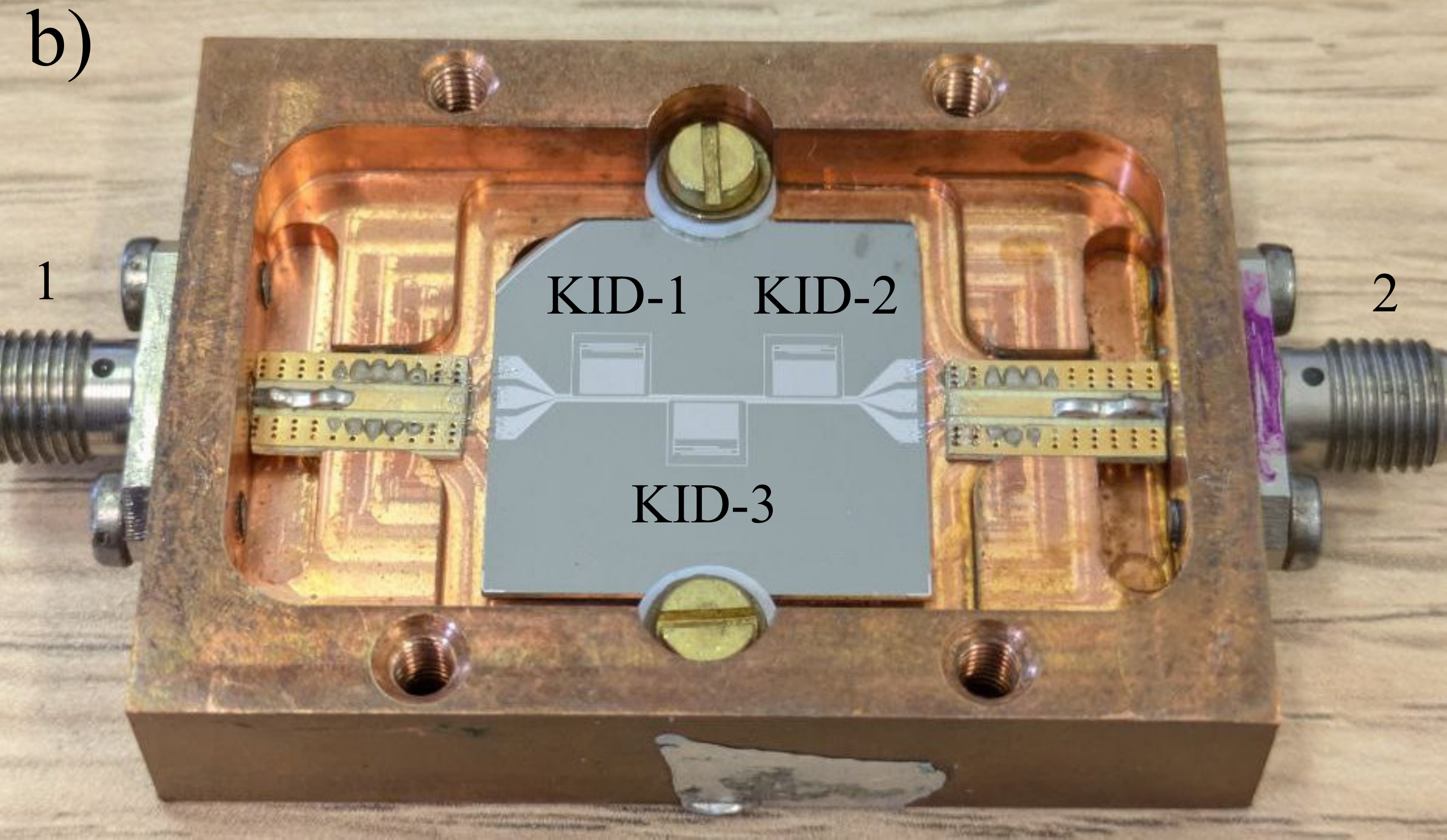}
\caption{
\textbf{(a)} Design of the single KID sensor. The layout of the resonator is identical to the ones of the BULLKID \cite{bullkid2022} silicon arrays: the 4~mm$^2$ inductor is unchanged while the coupling with the coplanar waveguide has been adapted to take into account the different dielectric constant of the germanium substrate.
\textbf{(b)} The 3 pixel array deposited on the 2x2~cm$^2$ germanium tile. The device is installed in a copper holder and held by Teflon supports. The KIDs are coupled to a single coplanar waveguide feedline for simultaneous readout via SMA connectors installed on the copper frame (ports 1 and 2).}
\label{fig1}
\end{figure}
The device has been fabricated using a 550~$\mu$m thick, 2" diameter  single side polished germanium wafer at the PTA clean room facility in Grenoble~\cite{salle-blanche}.
To fabricate the resonators the wafer is initially put under high vacuum and the native $\text{GeO}_2$ on the surface is removed by exposure to argon plasma. Afterwards the polished surface is coated by a 60~nm aluminium film by means of an electron gun evaporator. During the cleaning and coating of the wafer the vacuum is never broken, so that the oxide is not allowed to reform and contaminate the metal-substrate interface.

The removal of the germanium oxide is crucial to achieve a sufficiently high quality of the final device: \mycolor{a 10 minutes argon plasma etch achieved a complete removal of the germanium oxide, resulting in KIDs with quality factors and noise comparable to those on silicon substrates. Incomplete removal of the germanium oxide results in greatly reduced internal quality factors \mycolorbis{(e.g. $Q_i < 10$~k)}, as shown by prototypes which used a shorter argon plasma etching time of 3 minutes.}

After the argon etching process and the deposition of the metal layer, a standard process of UV lithography and wet etching is used to selectively remove the aluminium in order to form the structures of the resonators and of the coplanar waveguide to which the KIDs are inductively coupled. For redundancy the array is replicated four times on a single wafer that is then diced into 2x2~cm$^2$ square tiles.

The pixel design is identical to the sensors of the BULLKID array~\cite{bullkid2022}. 
The active area of each pixel is 4~mm$^2$, made of 105 strips of 1880$\times$20~$\mu$m$^2$ spaced by 8~$\mu$m and arranged to form an inductive meander. The two ends of the meander are terminated to four capacitive fingers whose length is \mycolor{adjusted} in order to select the resonant frequency of each pixel (Fig.~\ref{fig1} a); in the current version of the array three pixels are fabricated along a single 2~cm long coplanar waveguide. The sensors were simulated using the SONNET~\cite{sonnet} software: the capacitance has been chosen in order to have the devices resonate at frequencies $1/2\pi\sqrt{LC}$ between 0.7 and 0.8~GHz. The spacing in frequency between the resonators is 25~MHz to minimize electrical cross-talk. In order of increasing resonant frequency, KID-1 and KID-2 are on the same side of the feedline, while KID-3 is on the opposite side (Fig.~\ref{fig1} b). The coplanar waveguide geometry has been simulated and optimized for the dielectric constant of germanium $\epsilon_r^{Ge} \sim 16$ (while $\epsilon_r^{Si} \sim 12$)~\cite{ge_si_dielectric} in order to provide an impedance of $50~\Omega$.

%\section{EXPERIMENTAL SETUP}
\label{experimental_setup}
%rivedi la superficie di contatto!
The sample is installed in a copper holder held by Teflon supports and wire bonded to $50~\Omega$ launchers and then soldered to SMA connectors for readout, as shown in Fig.~\ref{fig1} (b). 
The contact area between the chip and the supports is reduced to $\sim 15$ mm$^2$ in order to minimize the phonon escape~\cite{martinez2019}.
The array is operated in a dry $^3\text{He}$/$^4\text{He}$ dilution refrigerator at a temperature $T = 220$~mK, which is well below the transition temperature of aluminium ($\sim$1.2~K).
The sample is protected by radiation from the warmer sections of the cryostat by means of thermal shields. Moreover the copper holder is wrapped in aluminium foil to further protect the array from stray light; finally a mu-metal barrel is placed outside of the cryostat to mitigate the effects of magnetic fields.
To calibrate the device the surface of the tile opposite to the KIDs is faced to an optical fiber placed between KID-1 and KID-2, pulsed by a 400~nm LED lamp operated at room temperature~\cite{Cardani_2018}.
The readout waves for the KIDs are generated at room temperature by an Ettus X310 board~\cite{ettus} and are routed to the array through a cryogenic transmission line with an attenuation of 55~dB to reduce thermal noise. The output signal from the array is then fed into a low-noise amplifier installed on the 4~K stage of the cryostat and demodulated at room temperature by the same board used for tone generation~\cite{minutolo}. 

%\section{DATA ANALYSIS AND RESULTS}
\begin{figure}[t]
\centering
\includegraphics[width=0.48\textwidth]{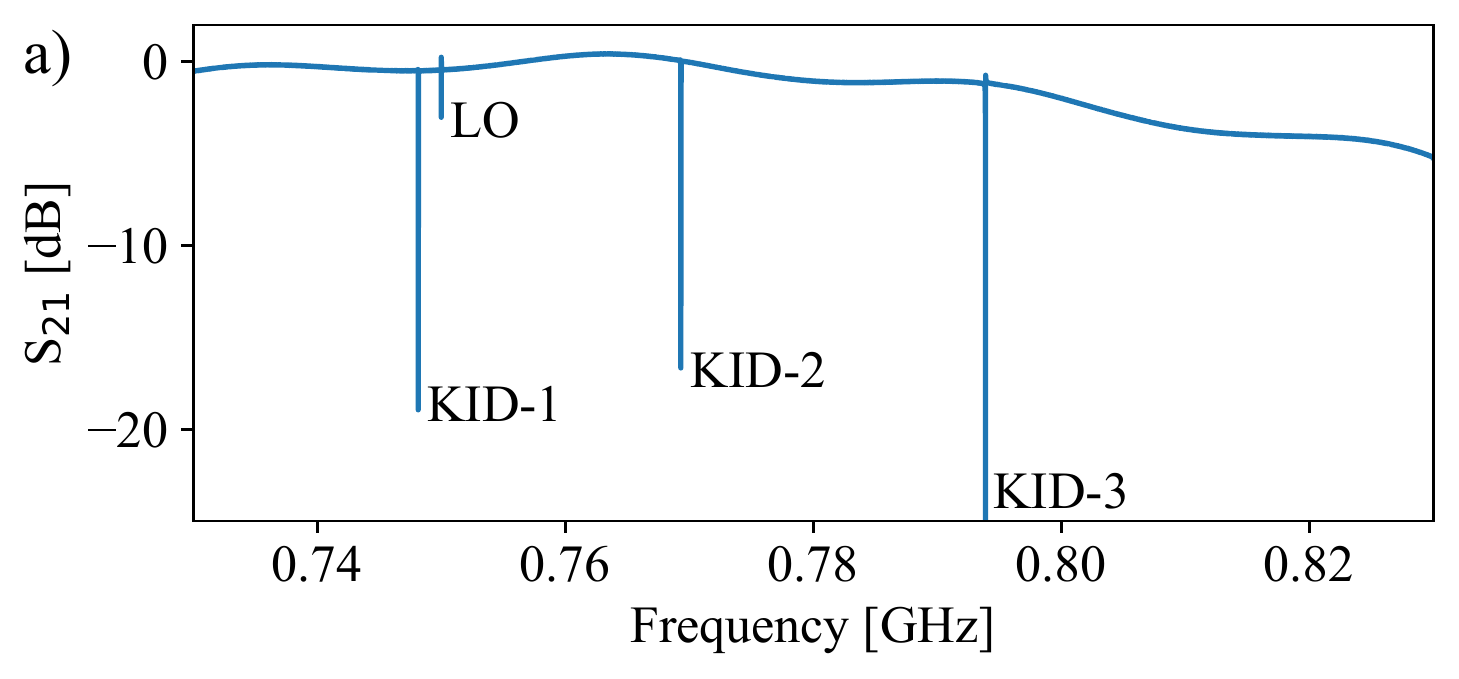}
\includegraphics[width=0.22\textwidth]{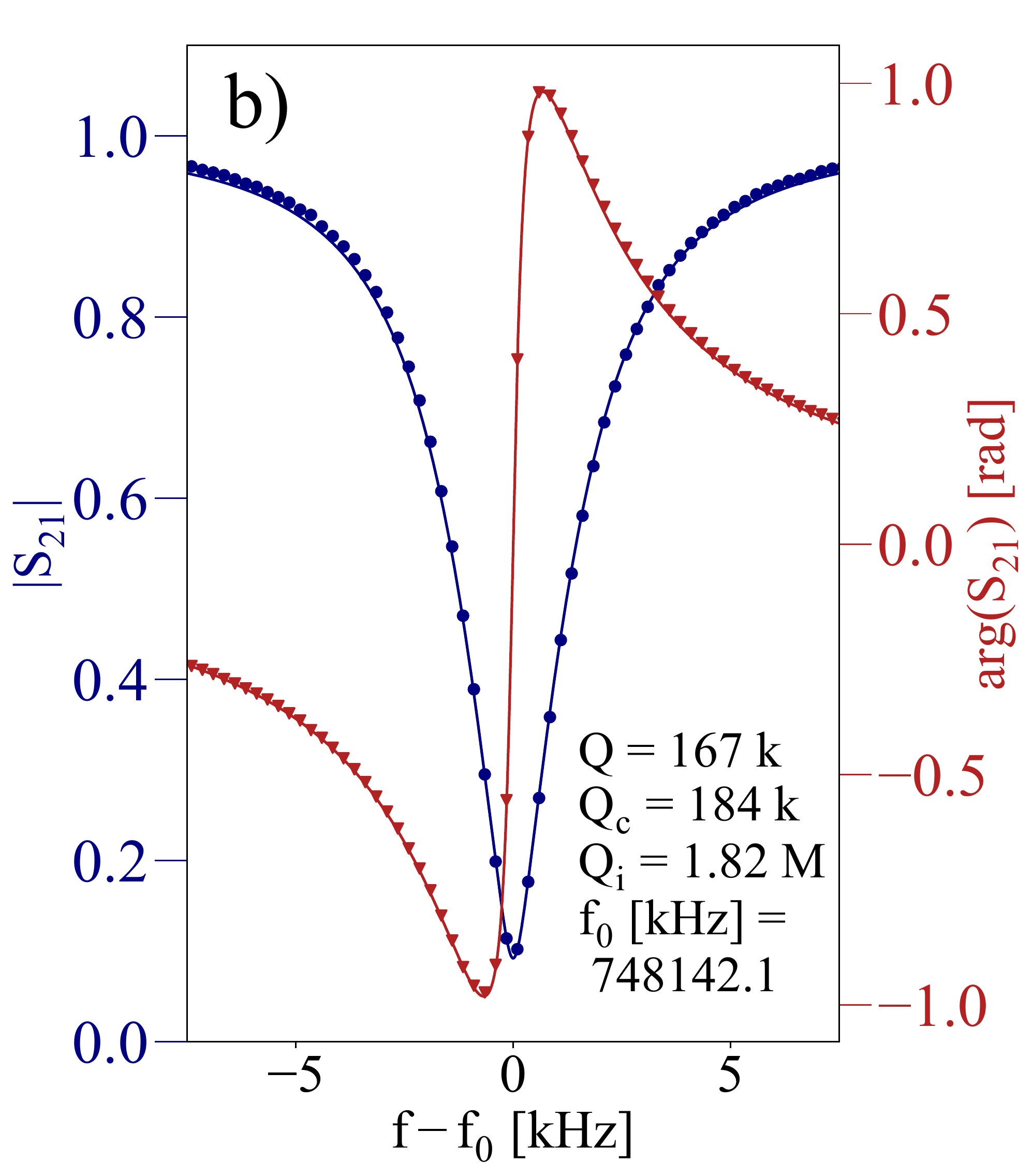}\includegraphics[width=0.27\textwidth]{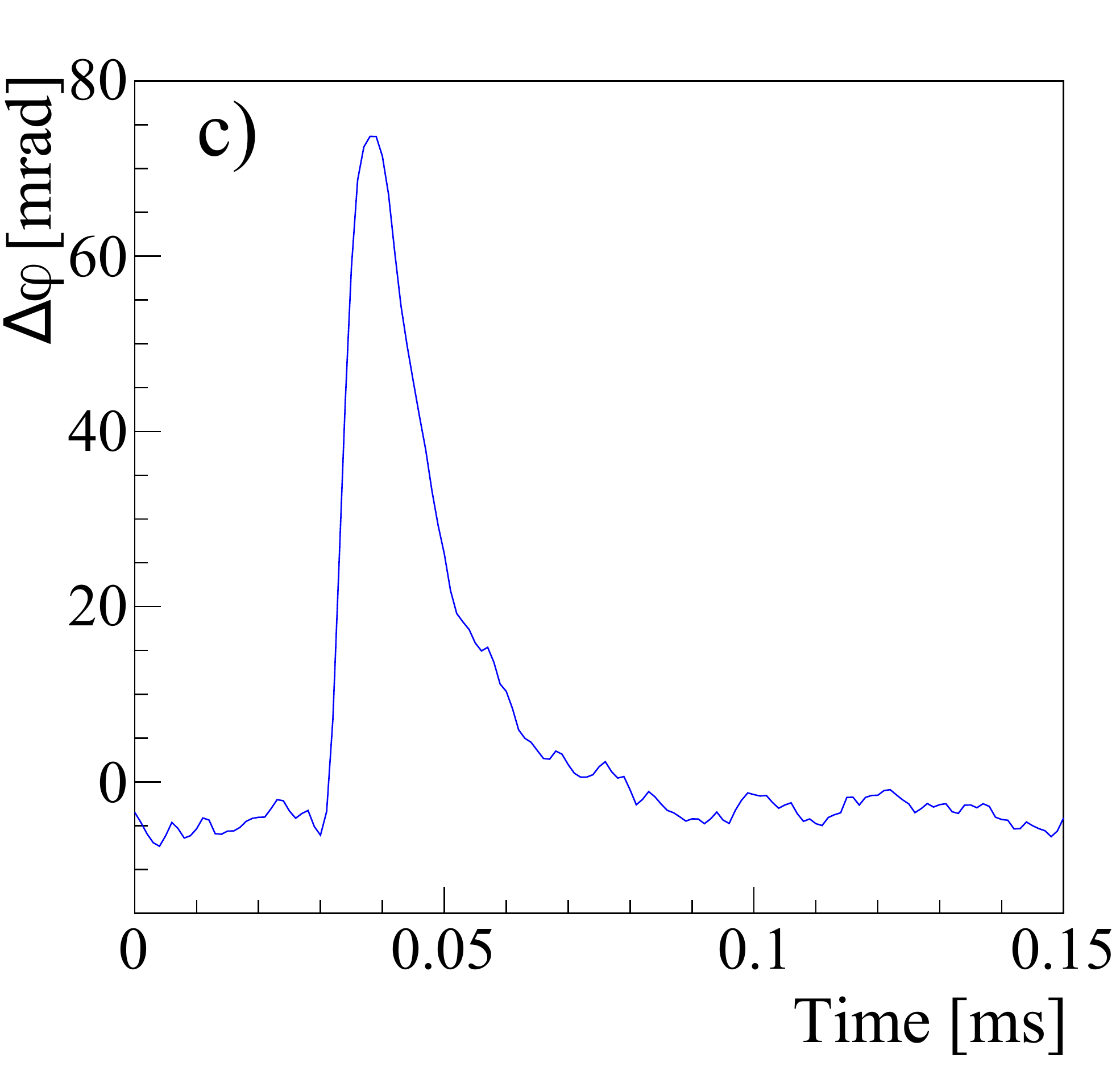}
\caption{
\textbf{(a)} Amplitude of the transmission $S_{21}$ of the array as a function of the readout frequency. The three minima correspond to the resonant frequency of each KID. The feature at 750~MHz is interference introduced by the local oscillator of the readout board (LO).
\textbf{(b)} Fit of the complex $S_{21}$ parameter of KID-1 to estimate the quality factors and the resonant frequency. 
\textbf{(c)} Phase pulse recorded by KID-1 after an energy deposition of approximately 20~keV induced by firing a LED pulse on the side opposite to the KIDs. The rise and decay times are 60~$\mu$s and 120~$\mu$s, respectively.}
\label{fig2}
\end{figure}
The measured transmission $S_{21}$ as a function of the frequency $f$ is reported in Fig.~\ref{fig2} (a). Each resonator is identified by a dip in the transmission and all of them are detected. The KIDs are spaced by $23\pm2$~MHz in accordance with the SONNET simulations previously discussed. 
In order to extrapolate the electrical properties of the resonators the data of the complex transmission $S_{21}(f)$ past each KID resonator are corrected for line losses and impedance mismatches~\cite{khalil} and fitted to the analytical expression~\cite{zmu_annrev2012}
\begin{equation}
    \label{S21}
    S_{21}(f) = 1 - \dfrac{Q}{Q_c} \dfrac{1}{1+j2Q\frac{f-f_0}{f_0}}
\end{equation}
where $f_0$ is the resonant frequency, $Q = (1/Q_c + 1/Q_i)^{-1}$ is the total quality factor, $Q_c$ is the coupling quality factor representing the coupling with the feedline and $Q_i$ is the internal quality factor representing internal losses.

The fit to Eq.~\ref{S21} to the $S_{21}$ data of KID-1 is shown in Fig.~\ref{fig2} (b), while a summary of the parameters of the whole array is reported in Tab.~\ref{tab1}. 
\setlength{\tabcolsep}{0.5em} 
\begin{table}[tb]

\centering
\begin{tabular}{lcccccc}

\hline
    & $f_0$ [MHz]   & $Q$ [k] \qquad & $Q_c$ [k] \qquad & $Q_i$ [M]\qquad \\ 
\hline
KID-1 & 748.1 &  167  &  184  &  1.82  \\
KID-2 & 769.3 &  114  &  120  &  2.24  \\
KID-3 & 793.9 &  127  &  132  &  3.03  \\
CALDER-17 & 2644 &  147  &  156  &  $>2$  \\

\hline
\end{tabular}
\caption{Results of the fit for Eq.~\ref{S21} of the data acquired from the frequency scan of the array. Data from a single KID on a silicon substrate are reported for comparison~\cite{Cardani:2017qr}.}
\label{tab1}
\end{table}
Since $Q_i \gg Q_c$ the value of the total quality factor $Q$ in the current configuration is dominated by $Q_c$: the measured value from Table~\ref{tab1} is in the range $Q =\left(114\text{ k} - 167\text{ k}\right)$, in line with the design value of $160$~k. The table also also reports the resonance parameters of a similar KID deposited on a silicon substrate (CALDER-17)~\cite{Cardani:2017qr}. Despite the different resonant frequency, which depends on the layout of the resonator, the quality factors are comparable.\\

In order to test the response of the array with respect to energy depositions, each KID is biased with an independent RF wave centered at the resonant frequency. After an energy deposition $E$ in the substrate phonons propagate until they reach the KIDs and break Cooper pairs. The change in Cooper pair density is detected by monitoring the variation of magnitude and phase of $S_{21}(f_0)$ as a function of time.
The phase readout usually provides a better signal to noise ratio~\cite{zmu_annrev2012} and for this reason we restrict the analysis to it.

The model of the phase response reads~\cite{MazinPhD}:
\begin{equation}
    \label{responsivity}
    \dfrac{d\phi}{dE} = \eta \cdot \dfrac{\alpha S_2(f_0,T_0)Q}{N_0 \Delta_0^2 V}
\end{equation}
where $N_0 = 1.72 \cdot 10^{10} eV^{-1}\mu$ m$^{-3}$ is the single spin density of states in aluminium, $V=4~\text{mm}^2\times 60$~nm is the inductor volume, $S_2(f_0,T) = 3.4$ is a dimensionless factor given by the Mattis–Bardeen theory related the imaginary part of the conductivity~\cite{zmu_annrev2012} and $\eta$ is the energy to quasi-particle signal conversion efficiency. The ratio of kinetic inductance over total inductance $\alpha = (4.7 \pm 0.1 )\%$ and the superconductor gap $\Delta_0 = 189 \pm 1$~$\mu$eV are estimated from the frequency shift of the resonators with respect to temperature~\cite{GAOvsMattisBardeen}.

The \mycolor{impulse} response of each KID has been characterized as a function of the readout power by firing pulses from the LED lamp at constant energy~\cite{Cardani:2017qr}. The pulses are processed offline with a matched filter in order to maximize the energy resolution~\cite{Radeka:1966}. The readout power optimizing the signal to noise ratio is -64~dBm at each KID. 

Figure~\ref{fig2} (c) shows a sample pulse corresponding to an energy deposition in the substrate of approximately 20~keV, as recorded by KID-1. 
The rise time amounts to $\sim 60~\mu$s, does not change with the temperature of the array and is comparable to the ring time of KID-1 $\tau_r = Q /\ \left( \pi f_0 \right) \approx 70$~$\mu$s.
%calculated as the time difference between 90\% and 30\% of the pulse height,
The decay time amounts to $120\pm 3$~$\mu$s and varies with the temperature, allowing to identify it as dominated by the quasi-particle recombination time $\tau_{qp}$~\cite{BarendsTau}.

\begin{figure}[t]
\centering
\includegraphics[width=0.5\textwidth]{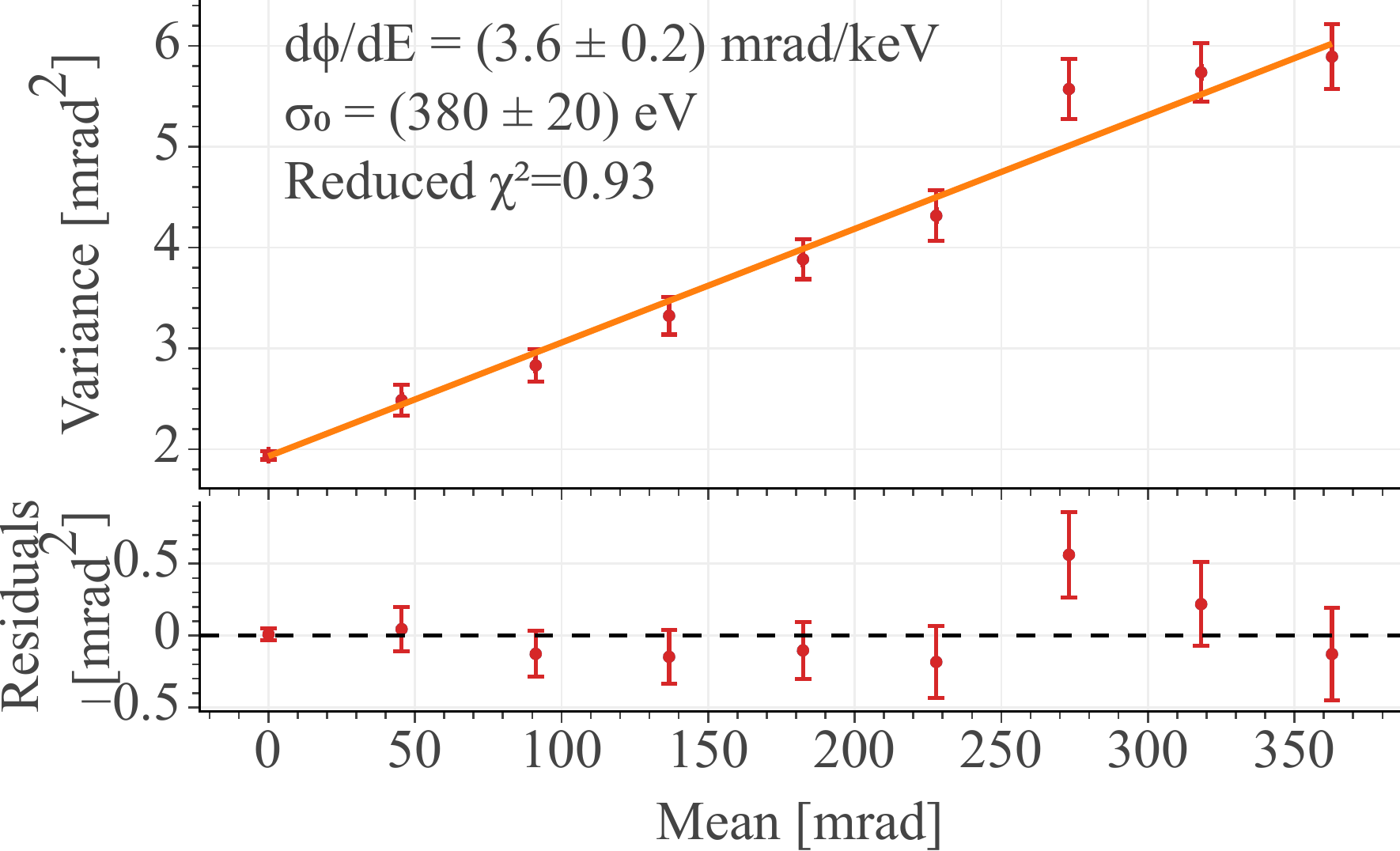}
\caption{Fit of $\sigma^2$ as a function of $\mu$ (Eq.~\ref{absolute-calibration}) corresponding to the response of KID-1 to bursts of 400~nm photons fired by the calibration LED. The linear fit converges to a responsivity of 3.6~mrad/keV and an energy resolution of 380~eV. This linear best fit is representative of the other units KID-2 and KID-3 whose calibration parameters are reported in Table~\ref{tab2}.}
\label{fig3}
\end{figure}

The estimation of the responsivity $d\phi/dE$ follows the procedure \mycolorbis{validated against a 6~keV X-ray source in Ref.~\cite{Cardani_2018} and employed in Refs.~\cite{Cardani:2021wl,bullkid2022,delCastello_2024}.}
The mean response of the detector to the absorption into the substrate of $N$ photons can be written as $\mu = d\phi/dE \cdot \epsilon N$ where $\epsilon = 3.1$~eV is the energy of a single 400~nm photon. 
The variance of the response is the sum of the variance of the noise $\sigma_0^2$ and the Poisson's variance:
\begin{equation}
    \label{absolute-calibration}
    \sigma^2(\mu) = \sigma_0^2 + \dfrac{d\phi}{dE} \cdot \epsilon \mu
\end{equation}
The energy calibration is performed by sending bursts of photons and evaluating $\sigma^2$ for different values of $\mu$. Fitting Eq.~\ref{absolute-calibration} for the parameters ${d\phi}/{dE}$ and $\sigma_0^2$ yields the responsivity in mrad/keV and the baseline energy resolution, respectively. An example of such estimation on KID-1, is depicted in Fig.~\ref{fig3}. The responsivity is $3.6\pm0.2$~mrad/keV and $\sigma_0$ is $380\pm20$~eV, while the results of the calibration for all of the KIDs are summarized in Tab.~\ref{tab2} and compared to CALDER-17.
\setlength{\tabcolsep}{0.25em} 
\begin{table}[tb]

\centering
\begin{tabular}{lcccccccc}

\hline
   & \makecell{ \mycolor{Tile} \\ \mycolor{mass $\left[\text{g}\right]$}} &  \makecell{ $d\phi/{dE}$ \\ $\left[\text{mrad/keV}\right]$} & $\sigma_0$ [eV] & $\eta$ [\%]   & $\dfrac{A_\text{KID}}{A_\text{TOT}}$\\ 
\hline
KID-1 & \mycolor{1.16}& $3.6\pm0.2$ &  $380 \pm 20 $  &  $2.0\pm0.2$ & 0.12   \\
KID-2  & \mycolor{"} & $2.5\pm0.1$ &  $450 \pm 22 $  &  $2.0\pm0.2$ &  "  \\
KID-3 & \mycolor{"} &$2.2\pm0.2$ &  $540 \pm 31 $  &  $1.6\pm0.2$ &  "  \\
CALDER-17 & \mycolor{0.28} & $5.8$ &  $115 \pm 6 $  &  $7.4-9.4$ & 0.42 \\
\hline
\end{tabular}
\caption{Results of the fit of Eq.~\ref{absolute-calibration} to the data acquired by firing LED pulses on the backside of the array. The value of the phonon energy conversion efficiency $\eta$ are derived by inverting Eq.~\ref{absolute-calibration}. The efficiency is proportional to the ratio of sensitive metal area to the total metal area $A_\text{KID}/A_\text{TOT}$. \mycolor{Values for CALDER-17 are also reported for comparison, as well as the total mass of the Ge and Si substrates.}}
\label{tab2}
\end{table}

By inverting Eq.~\ref{responsivity} we extract the energy to quasi-particle conversion efficiency $\eta$. The resulting values are summarized in Tab.~\ref{tab2}, in the range $(1.6 - 2.0)\%$. Since the optical fiber is firing on a spot between KID-1 and KID-2 it is expected that they share a similar value while the more distant KID-3 has a reduced efficiency. 
The energy conversion efficiency is a factor 4 lower than in the CALDER-17 prototype that achieved $(7.4-9.4)\%$~\cite{Cardani_2018} depending on the position of the fiber spot. 

\mycolorbis{The origin of the reduced value of $\eta$ can be understood by comparing the layout of the two devices. The CALDER-17 detector implements a single KID, so that phonons produced can either escape via the substrate supports, be absorbed by the KID or by the insensitive metal (ie. the feedline and the capacitor). In the device of this work, aside the supports and the insensitive metal, the three KIDs share the phonons~\cite{martinez2019}.
The ratio of the active metal surface of each KID over the total metallized surface (${A_\text{KID}}/{A_\text{TOT}}$) is approximately $0.42$ in CALDER-17, while it is only $0.12$ for one of the KIDs of this work. Moreover, a single-side polished wafer was used for the realization of the current array, in place of the double-sided polished wafer for CALDER-17.} 
It is reasonable to believe that an additional source of loss is given by the unpolished surface that could act as a trapping site for phonons.

\mycolorbis{We conclude that there} is no evidence that the material used as substrate influences the responsivity and the energy resolution. \mycolorbis{This seems to be backed up also by the similar values of the expected athermal phonon transmission to Aluminum from Germanium and Silicon. Following the method in Ref.~\cite{Kaplan1976}, we indeed obtain $\sim90\%$ and $\sim80\%$ phonon transmission to Aluminum from Silicon and Germanium, respectively.}

We can therefore ascribe the difference in baseline resolution with respect to CALDER-17 only to the geometry of the metal layer on the tile, which affects the phonon collection efficiency. \mycolorbis{If we assume the value of $\eta=24\%$ of a BULLKID-like geometry, we expect an improvement of the baseline energy resolution of KID-1 down to $380\times2\%/24\%=32$~eV, in line with the 26~eV of the silicon prototype.}

%%%%%%%%%%%%%%%%%%%%%%%%%%%%%%%%%%%%%%%%%%%%%%%%%%%%%%%%%%%
%\section{CONCLUSION and PERSPECTIVES}
In this work we presented the design, fabrication and operation of the first \mycolorbis{KIDs developed on a germanium substrate.} The fabrication process was optimized to remove the germanium dioxide from the substrate, as it was found to be a limiting factor for the quality of the resonators. We estimated detection performances which are in line to that obtained with silicon substrates, paving the way to future developments for applications in the neutrino and dark matter fields.

\begin{acknowledgments}
We thank the team of the Laboratorio Rivelatori Criogenici of Sapienza U. for its support. This work was partially supported through the European Research Council through the Consolidator Grant “DANAE” number 101087663.
We acknowledge the support of the PTA platform for the fabrication of the device. We thank A. Girardi and M. Iannone of the INFN Sezione di Roma for technical support. \mycolor{We thank A. L. De Santis for useful discussions on phonon propagation.}
\end{acknowledgments}

\section*{Data Availability Statement}
The data that support the findings of this study are available from the corresponding author upon reasonable request.

\bibliographystyle{apsrev4-1}
\bibliography{calder_apl}

\end{document}